# Structural and magnetic properties of epitaxial films of CoIrMnAl equiatomic quaternary Heusler alloy designed from first-principles calculation

Ren Monma[a,b], Tufan Roy[c], Kazuya Suzuki[b,d], Tomoki Tsuchiya[e,d], Masahito Tsujikawa[c,d], Shigemi Mizukami[b,d,e], Masafumi Shirai[c,d,e]

[a]Department of Applied Physics, Graduate School of Engineering, Tohoku University, Aoba 06-05, Sendai 980-8579, Japan

[b]WPI Advanced Institute for Materials Research (AIMR), Tohoku University, 2-1-1, Katahira, Sendai 980-8577, Japan

[c]Research Institute of Electrical Communication, Tohoku University, Sendai 980-8577, Japan

[d]Center for Spintronics Research Network (CSRN), Tohoku University, Sendai 980-8577, Japan

[e]Center for Science and Innovation in Spintronics (CSIS), Core Research Cluster (CRC), Tohoku University, Sendai 980-8577, Japan

**Abstract**

MgO-barrier magnetic tunnel junctions with half-metallic Heusler alloy electrodes attracted much attentions for spintronics applications. However, a couples of issues related to materials still remain to be resolved for practical uses. Recently, quarterly equiatomic Heusler alloys attracted attentions as advanced Heusler alloys. CoIrMnZ (Z = Al, Si, Ga, and Ge) half-metallic Heusler alloys were designed and predicted to have moderate Curie temperatures and to be a lattice-matched with the MgO barrier, being advantageous to traditional $Co_2$ Heusler alloys [T. Roy *et al.,* J. Magn. Magn. Mater. 498, 166092 (2020)]. Here we experimentally investigated structure and magnetic properties for thin films of one of those alloys, CoIrMnAl with a sputtering deposition. We successfully obtained the films with the *B*2 chemical ordering even with no post-annealing process. The lattice constant for the films annealed at 500–600°C approximates the predicted values. The magnetization at 10 K was near 500 kA/m and the Curie temperature was approximately 400 K were observed, which were about 70% of the values predicted for the fully ordered structure. The magnetic properties observed in those *B*2 ordered films were well explained by ferrimagnetism appeared in *B*2 ordered CoIrMnAl with full-swap disorders of Co-Ir and Mn-Al and almost full-swap disorder of Co-Mn, predicted from the first-principles calculations.

## 1. Introduction

A magnetic tunnel junction (MTJ) exhibits tunnel magnetoresistance (TMR) effect [1, 2, 3], and is utilized in various storage system [4, 5] and non-volatile magnetoresistive random access memories (MRAM) [6]. More recently, advanced computing technologies are emergent and demand MTJs exhibiting huge TMR ratio at room temperature [7, 8, 9]. Huge TMR ratios can be obtained in MTJs with uses of electrodes of half-metallic ferromagnets which are magnetic metals having a band gap at the Fermi level for their minority spin states [10, 11]. A $Co_2$-based full Heusler alloy is one of the candidates of half-metals and attracted much attention to date [12, 13]. Several $Co_2$-based full Heusler alloys have been utilized to the electrodes of the MTJs [14, 15, 16, 17, 18]. Among them, a promising alloy for the electrodes of MTJs is $Co_2MnSi$ and its derivative. Indeed, MTJs with electrodes of $Co_2MnSi$ or of the ones with Mn partially substituted by Fe showed the record of the TMR ratio exceeding 2600% at low temperature [17]. However, this TMR ratio is significantly reduced at room temperature [17, 19]. The reduction of the TMR ratio with elevating temperature may be due to the reduction of the spin polarization and thermal fluctuation of magnetic moment at their hetero-interface of MgO(001) barrier/Heusler alloys(001) [19, 20, 21, 22, 23]. In addition, there are many discussions on intrinsic and/or extrinsic interface electronic states existing at those interfaces reducing the TMR ratio [24, 25]. Thus, it is of crucial to find unique Heusler alloys having their lattice constant matching to that of MgO(001) as well as unique interface electronic structures robust against termination elements.

Such candidates may be found in equiatomic quaternary Heusler alloys with a chemical formula of XX'YZ [26, 27, 28], where X, X', and Y denote transition metal elements and Z represents a main group element. Fully ordered crystal structure is a cubic LiMgPdSn or $Y$ type, as shown in Fig. 1(a). X', X, Z, and Y occupy at the Wyckoff positions, $4a$ (0, 0, 0), $4b$ (1/2, 1/2, 1/2), $4c$ (1/4, 1/4, 1/4), and $4d$ (3/4, 3/4, 3/4), respectively [28]. In recent years, intensive theoretical and experimental studies on numerous elemental compositions in those alloys and many interesting properties were discussed [29, 30, 31, 32, 33, 34, 35, 36, 37, 38]. A couples of those alloys were examined as the electrode of the MTJs [39, 40]. Equiatomic quaternary Heusler alloys could have various chemically disordered states, *e.g., XA* [Fig. 1(b)], $L2_1$ (or $L2_1B$) [Fig. 1(c)], $B2$ [Fig. 1(d)], and $A2$ [Fig. 1(e)], which belong to different space groups. Therefore, it is also crucial to understand the influence of such disordered states to the various properties [40, 41, 42, 43, 44].

The part of the present authors have theoretically studied equiatomic quaternary Heusler alloys CoIrMnZ (Z=Al, Si, Ga, Ge) [37]. The ground state for those alloys are ferromagnetic and those Curie temperatures $T_C$ are well above room temperature. Spin polarizations were close to unity owing to the band gap at the Fermi level in their minority spin states. Their cubic lattice constants are relatively larger than that of Co$_2$MnSi and their lattice mismatches to MgO(001) are quite small, *e.g.,* 0.81% for Z=Al; thus one anticipates to obtain better interface of CoIrMnZ(001)/MgO(001) MTJs [37].

In this article, we experimentally investigated, for the first time, structure and magnetic properties for CoIrMnAl thin films. We successfully obtained $B2$ ordered CoIrMnAl epitaxial films even without any heat treatments with a sputtering deposition technique. Saturation magnetization at 10 K was near 500 kA/m and the Curie temperature was approximately 400 K. Those values were approximately 70% of the values predicted for $Y$ ordered CoIrMnAl [37]. We discuss different types of disorders and effects of those on magnetic properties using the first-principles calculation. The magnetic properties observed in our films are well explained by ferrimagnetism in $B2$ ordered CoIrMnAl when we take account of full-swap disorders of Co-Ir and Mn-Al and almost full-swap disorder of Co-Mn.

## 2. Experimental procedures and first-principles calculation details

All films were deposited on single crystalline MgO(100) substrates using the magnetron sputtering using Co, Ir, and MnAl alloy targets. A base pressure is lower than $2\times10^{-7}$ Pa. Prior to the deposition of the films, the MgO substrates were flushed at approximately 700°C in the chamber. The 50-nm-thick

CoIrMnAl films were deposited at room temperature followed by an in-situ post annealing with an annealing temperature $T_a$ of 300, 400, 500, and 600°C. All films were capped by the 3-nm-thick Ir after cooling the substrate. The composition for the films (at%) is Co : Ir : Mn : Al = 28.1 : 27.6 : 20.7 : 23.6, which was analyzed with an inductively-coupled plasma mass spectroscopy (ICP-MS). Film structures were characterized by X-ray diffraction (XRD) using Cu $K_\alpha$ radiation. Magnetization of the films were measured using a vibrating sample magnetometer (VSM). Temperature dependence of magnetic properties were also measured by VSM in a physical property measurement system (PPMS) with temperature ranging from 10 to 400 K with an in-plane magnetic field application.

First-principles calculations were carried out using the spin-polarized-relativistic Korringa-Kohn-Rostoker (SPRKKR) method, as implemented in the SPR-KKR program package [45]. The calculations were performed in full-potential mode. We considered the substitutional disorder in the system within coherent potential approximation. Generalized gradient approximation was used for the exchange correlation functional as modelled by Perdew, Burke, and Ernzerhof (PBE) [46]. For all atoms, the angular momentum cut-off $l_{max}$ was restricted to two and the Brillouin zone was sampled with a $k$-mesh of dimension 26×26×26. For the determination Fermi energy, Lloyd's formula was used [47, 48]. As for magnetic interactions, we calculate the Heisenberg exchange coupling constant within a real space approach as proposed by Liechenstein *et al* [49]. Furthermore, we evaluate the Curie temperature $T_C$ in terms of the Heisenberg exchange coupling constant, within a mean-field approximation.

## 3. Experimental results

Figure 2 shows out-of-plane XRD patterns for the films with different post-annealing temperatures $T_a$. We observed the (002) superlattice and the (004) fundamental diffraction peaks for all films. Those (002) and (004) peaks shift to higher angle as the annealing temperature increases. However, the diffraction peaks from the (111) plane which should be observed in *Y*, *XA*, *L*2$_1$, and *D*0$_3$ type ordering were not detected in any films. This suggests that these films have *B*2 chemical ordering. It should be noted that the films exhibit the superlattice (002) peaks even though the films are not annealed.

Figure 3(a) shows the lattice constant *a* evaluated from the (004) diffraction peak. The lattice constant *a* is almost constant for the annealing temperature below 400°C and those decrease with increasing the annealing temperature above 400°C. The *a* values at 500–600°C are 0.596–0.597 nm, and those are relatively close to the theoretical value, 0.5905 nm, for *Y* ordered CoIrMnAl predicted [37]. Figure 3(b) shows the ratio of the integrated intensity for the (002) and (004) diffraction peaks for the films and those values tend to increase with increasing the post-annealing temperature $T_a$. The changed behavior in the integrated intensity ratio seems to correlate with those in the lattice constant and it shows relatively large increase above 400°C. The increase in the integrated intensity ratio indicates the increase in the long range $B2$ chemical ordering in our films. Hence, the films may have better $B2$ chemical ordering with $T_a =$ 500–600°C.

As shown in Fig. 1(d), $B2$ ordered state results from both the full-swap disorders of Co-Ir and Mn-Al, for example. Other types of disorders can also change, in principle, $Y$ ordered state into $B2$ state in equiatomic quaternary alloys because of multiple elements. Here, we gain more insight into types of disorders present in our films at $T_a =$ 500–600°C by comparing the experimental intensity ratio to the theoretical one. The theoretical peak intensity can be calculated with the expression [50]:

$$I = Cm|F|^2 LPA e^{-2M} \tag{1}$$

Here, $m$, $F$, $LPA$, and $M$ are the multiplicity, the structure factor, the Lorentz-polarization-absorption factor, and the Debye-Waller factor, respectively, and those depend on the diffraction peaks index [50]. $C$ is the constant independent from the diffraction peak index. We calculate $F$ for $B2$ ordered CoIrMnAl as the reduced $B2$ unit cell with the corner-site A and body-centered-site B occupied by four element atoms [Inset in Fig. 3(c)]. We neglect small off-stoichiometry for the films, for simplicity. The structure factor for the superlattice peak $F_s$ depends on the site occupancies of the four element atoms and it can be expressed as [50],

$$F_s = 2(f_A - f_B), \tag{2}$$

where the atomic form factor $f_{A(B)}$ is expressed with the weighted average of the atomic form factors for the atoms occupying at A (B) site:

$$f_A = f_{Co} x_{Co} + f_{Ir} x_{Ir} + f_{Mn}(1/4 - x_{Mn}) + f_{Al}(1/4 - x_{Al}), \tag{3}$$

$$f_B = f_{Mn} x_{Mn} + f_{Al} x_{Al} + f_{Co}(1/4 - x_{Co}) + f_{Ir}(1/4 - x_{Ir}). \tag{4}$$

Here, $f_j$ and $x_j$ are the atomic form factor and the occupation number at the correct site for the four element atoms, $j =$ Co, Ir, Mn, and Al, respectively. The value for $x_j$ ranges from 1/8 to 1/4. In case of $x_j =$ 1/4 for all $j$, Co and Ir (Mn and Al) occupy at A (B) site. Contrary, the case of $x_j =$ 1/8 for all $j$ represents $A2$ disordered state and all four element atoms randomly occupy at both A and B site. The structure factor for the fundamental peak $F_f$ is unchanged with chemical ordering:

$$F_f = 2(f_A + f_B) = (f_{Co} + f_{Ir} + f_{Mn} + f_{Al})/2 \tag{5}$$

Figure 3(c) shows the theoretical intensity ratio of the (002) and (004) diffraction peaks with different occupation numbers $x_j$. The theoretical values are found to be sensitive to the disorder of Ir and Al because of huge difference between the largest and smallest atomic form factor, $f_{Ir}$ and $f_{Al}$, respectively. In case that $x_{Ir}$ is varied with $x_{Co} = x_{Mn} = x_{Al} = 1/8$ fixed, i.e., only the Ir atom ordering, the intensity ratio shows large increase with increasing $x_{Ir}$ [solid circles in Fig. 3(c)]. Similarly, if $x_{Al}$ is varied with full disordering of the other elements, $x_{Ir} = x_{Co} = x_{Mn} = 1/8$, the intensity ratio also exhibits visible change [solid triangle in Fig. 3(c)]. The theoretical intensity ratio may become similar to the experimental values at $T_a$ = 500–600°C for the case shown with the solid squares in Fig. 3(c), at $x_{Ir} = x_{Al} = 1/4$. This is the case that Ir and Al fully occupy at A and B sites, respectively, while Co and Mn randomly occupy. These values are not much sensitive to the ordering of Co and Mn atoms and are nearly the same even in case of Co (Mn) occupying at A (B) site because of $f_{Co} \sim f_{Mn}$. Therefore, the XRD data shown in Fig. 1(b) suggests that our films annealed at 500-600°C are close to $B2$ ordered state with Ir and Al separately occupying A and B site, whereas the occupancies of Co and Mn are not clear from those XRD data.

Figure 4(a) shows the magnetization hysteresis loops for the films annealed at different post-annealing temperatures $T_a$. These measurements were carried out with the in-plane magnetic field applications. Figure 4(b) displays $T_a$ dependence of the saturation magnetization $M_s$ at room temperature. The values of $M_s$ lie in between 300-350 kA/m at $T_a$ = 500–600°C, probably correlated to the change in the structures observed in the XRD data [Figs. 3(a) and 3(b)].

Figure 5(a) shows the magnetization hysteresis loop measured with different temperatures $T$ for the film annealed at 500°C. For this measurement, we reprepared the film with less concentration of Co and Ir and the composition closer to the stoichiometry. Temperature dependence of the saturation magnetization $M_s$ is also shown in Fig. 5(b). The $M_s$ value at 10 K is close to 500 kA/m, and this value was by a factor of approximately 0.7 smaller than the theoretical value, 722 kA/m (4.03 $\mu_B$/f.u.) [37]. Our measurement set-up only allows us to raise temperatures up to 400 K and the precise value of the Curie temperature $T_C$ for the film is not obtained. Hence, the theoretical curve for $M_s$ vs $T$ was calculated and shown in Fig. 5(b) using the approximate equation [51]:

$$M_S = M_{s0}\left[1 - \left(\frac{T}{T_C}\right)^2\right]^{\frac{1}{2}}. \tag{6}$$

The experimental data is fitted to the data calculated with $M_{s0}$ = 485 kA/m and $T_C$ = 405 K. Thus we consider $T_C$ for the film is approximately 400 K, which is also by a factor of approximately 0.7 smaller than the predicted values, 584 K, for the fully ordered structure [37].

## 4. Discussions

Chemical disorders in our films may be one possibility of the origin for the difference in magnetization $M_{s0}$ and the Curie temperature $T_C$ between the experiment and the prediction for $Y$ ordered CoIrMnAl [37]. For further understanding, we performed the first-principles calculation of the electronic structures for CoIrMnAl with various atomic disorders. Figure 6 displays the formation energy $E_{Form}$ for full-swap disorders of various constituent atoms with respect to that of $Y$ ordered CoIrMnAl. The formation energy for the swap disorder of Mn-Al is the lowest among the disordered structures considered here, indicating that $Y$ ordered CoIrMnAl tends to have random swapping of Mn and Al and to change its ordered state from $Y$ to $L2_1$ [Fig. 1(c)]. This tendency is consistent with the theoretical results obtained for $Co_2$-based Heusler alloys [52]. Co-Mn and Co-Ir swap disorder show second and third lowest formation energies, respectively, and the difference of $E_{Form}$ between those two are very small. Hence, those Co-Mn and Co-Ir swap disorders could simultaneously occurs if those are introduced with some mechanism in addition to Mn-Al swap disorder. These swap disorders of both Co-Mn and Co-Ir in addition of Mn-Al swap disorder change the chemical order of CoIrMnAl from $L2_1$ to $B2$. On the other hand, the swap disorders of Co-Al, Ir-Al, and Ir-Mn exhibit relatively high formation energy, as show in Fig. 6, implying that Al and Ir tend to separately occupy at their own sites.

These insights are basically consistent with our analysis of the chemical ordering based on the XRD data (Fig. 3), whereas the degree of the swap disorder of Co-Mn were not clear from the experiments, as mentioned earlier. For further understanding, we also performed the first-principle calculation of the electronic structures with various amounts of the Co-Mn swap disorder. Figures 7(a) and 7(b) show the formation energy $E_{Form}$ and the total magnetic moments $M_{tot}$ for $B2$ CoIrMnAl with different degree of swap disorder of CoMn. Here, the disorder is represented by the occupation number of Co (Mn), $x_{Co(Mn)}$, at A (B) site, in $B2$ unit cell, as defined earlier in the inset in Fig. 3(c). And we assumed that Ir and Al occupy A and B site, respectively, *i.e.*, $x_{Ir} = x_{Al} = 1/4$ since the formation energy of Ir-Al swap disorder is much higher than that of Co-Mn disorder (Fig. 6). Regarding this calculation, we have considered two cases; the parallel and anti-parallel alignment of spins for Mn at A site (antisite) and at B site (ordinary site), which are denoted with $Mn^\uparrow_A$–$Mn^\uparrow_B$ and $Mn^\downarrow_A$–$Mn^\uparrow_B$, respectively, in Fig. 7. As seen in Fig. 7(a), the formation energies in the anti-parallel spin state ($Mn^\downarrow_A$–$Mn^\uparrow_B$) are smaller than in the parallel spin state ($Mn^\uparrow_A$–$Mn^\uparrow_B$) in almost entire range of $1/8 < x_{Co(Mn)} < 1/4$. Therefore, the $B2$ alloys assumed here may tend to be ferrimagnetic rather than ferromagnetic. This difference in magnetic ordering is clearly seen in Fig. 7(b), where such ferrimagnetic alloys show significant reduction of the total magnetic moment. The experimental magnetization measured at low temperature is close to the theoretical data at $x_{Mn}$ of about 0.14-0.15 in Fig. 7(b). We also calculated $T_C$ with different $x_{Co(Mn)}$, as shown in Fig. 7(c). The theoretical values of $T_C$ slowly decay with decreasing $x_{Co(Mn)}$, and are comparable to the experimental data at $x_{Co(Mn)}$ of about 0.15. Hence, the magnetic properties are well explained with the first-principles calculation at $x_{Co(Mn)}$ of 0.14-0.15. This $x_{Co(Mn)}$ value is close to $x_{Co(Mn)} = 0.125$, the full-swap disorder of Co and Mn, and this may be theoretically accord with the presence of the full-swap disorder of Co-Ir in our films, because those formation energies are comparable, as seen in Fig. 6.

Finally, we note that the swap disorders of not only Co-Mn but also Co-Ir and Mn-Al showed no reductions in magnetic moments of $Y$ ordered state when we assumed ferromagnetic spin alignments for all the elements. Hence, the reduction of the magnetic moment in the experiments is explained by ferrimagnetism owing to the swap disorder of Mn elements from the theoretical points of view. Meanwhile, our calculation indicated that the anti-parallel alignment of spins for Mn within B sites, which can be accompanied with the swap disorder of Mn-Al, is energetically very high. Thus, most realistic interpretation is the ferrimagnetic ordering accompanied with the swap disorder of Co-Mn, as discussed above.

## 5. Summary

We experimentally studied structural and magnetic properties for thin films of CoIrMnAl Heusler alloy, previously designed for spintronics applications. We successfully obtained the films with the $B2$ chemical ordering using the sputtering deposition technique even with no heat treatments. The lattice constants for the films annealed at 500-600°C were close to the predicted values. The XRD analysis for those films suggested that Ir and Al almost occupied at two different crystallographic sites in $B2$ structure. The saturation magnetization at low temperature, ~ 500 kA/m, and the Curie temperature of ~ 400 K were obtained and those were approximately 70% of those values predicted from the values in $Y$ ordered CoIrMnAl. The first-principles calculation were performed with taking account of the various swap disorders, and the experimental magnetization and the Curie temperature agreed well with the values calculated in $B2$ ordered CoIrMnAl with nearly full-swap Co-Mn disorders as well as full-swap Co-Ir and Mn-Al disorders. The predicted disordered state was also accord with the preference disorders in terms of their formation energies evaluated from the first-principle calculations. It will be a future subject to understand the physics and chemistry inducing such swap disorders and to find a way to control them.

**Acknowledgments**

We would like to thank Y. Kondo and K. Saito for their assistances and Kelvin Elphick and Atsufumi Hirohata for valuable discussions. This work was partially supported by JST CREST (No. JPMJCR17J5).

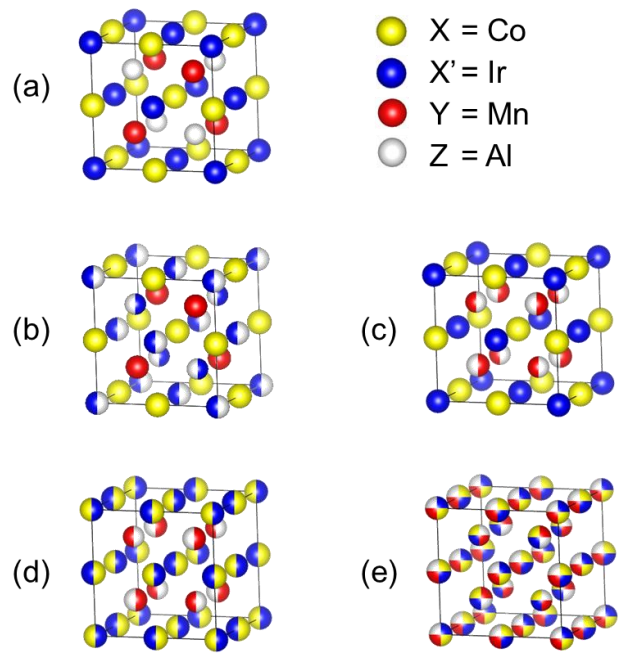

Figure 1: Schematics of various crystal structures of CoIrMnAl. (a) fully-ordered structure, *Y*. (b) *XA*, *e.g.*, with full-swap disorder of Ir-Al. (c) *L2$_1$* (or *L2$_1$B*), *e.g.,* with full-swap disorder of Mn-Al. (d) *B2*, *e.g.*, with both full-swap disorders of Co-Ir and Mn-Al. (e) fully-disordered structure, *A2*. *DO$_3$* ordered state is also possible with some full-swap disorders but is omitted here.

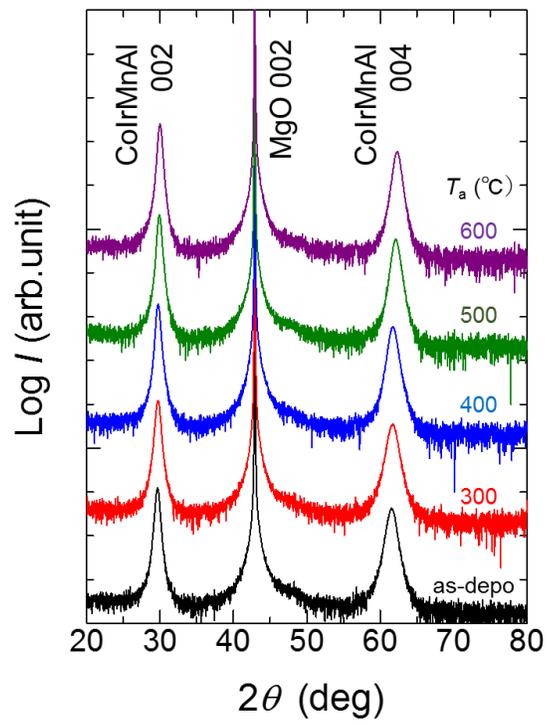

Figure 2: Out-of-plane 2$\theta$–$\omega$ XRD patterns for the films with different post-annealing temperatures $T_a$.

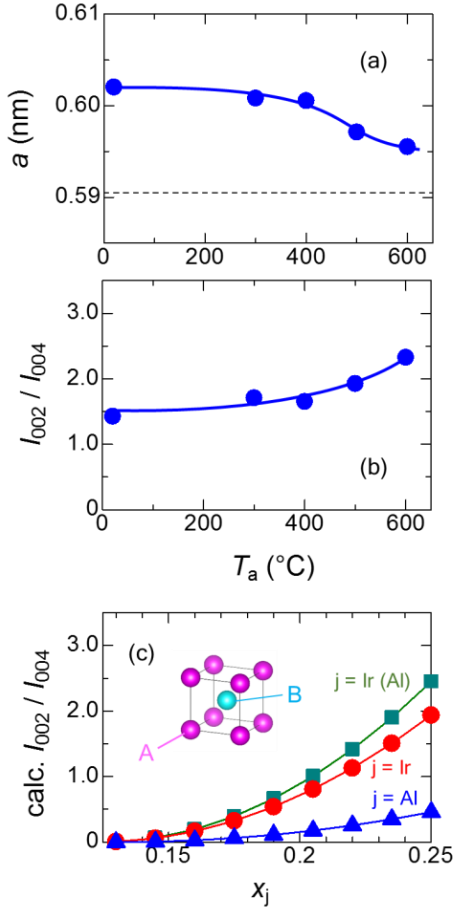

Figure 3: (a) The out-of-plane lattice constant for the films as a function of the post-annealing temperature $T_a$. The dashed line is the value predicted from the first principle calculations for $Y$ ordered CoIrMnAl [37]. (b) The integrated intensity ratio of the (002) and (004) peaks for the films as a function of the post-annealing temperature $T_a$. (c) The theoretical intensity ratio of the (002) and (004) peaks with different occupation numbers $x_j$. The case that $x_{Ir}$ is varied with keeping $x_{Co} = x_{Mn} = x_{Al} = 1/8$ (solid circles). The case that $x_{Al}$ is varied with keeping $x_{Ir} = x_{Co} = x_{Mn} = 1/8$ (solid triangle). The case that $x_{Ir}$ (= $x_{Al}$) is varied with keeping $x_{Co} = x_{Mn} = 1/8$ (solid square). Curves are visual guide. The $B2$ cell assumed in the calculation is in inset.

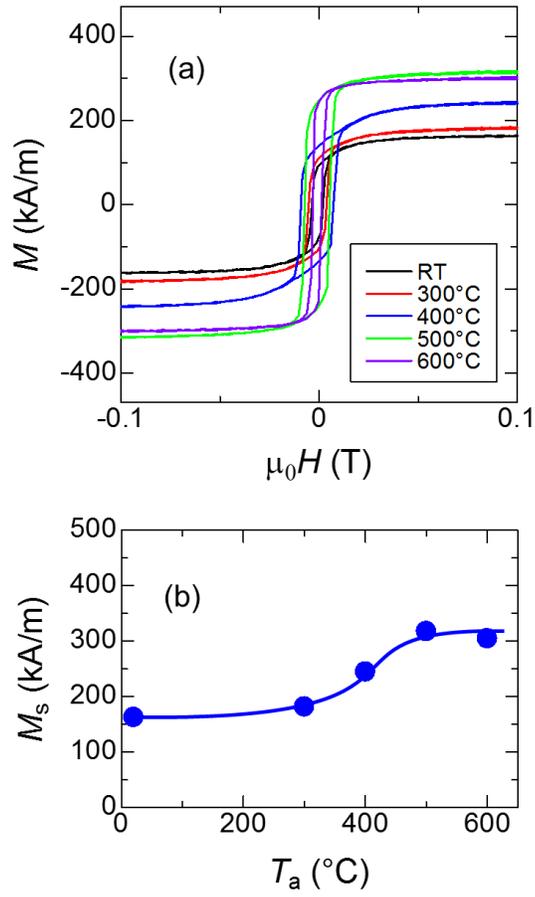

Figure 4: (a) Hysteresis loops of the magnetization $M$ for the films with different post-annealing temperatures $T_a$. (b) Saturation magnetization $M_s$ at room temperature as a function of $T_a$. Curve is the visual guide.

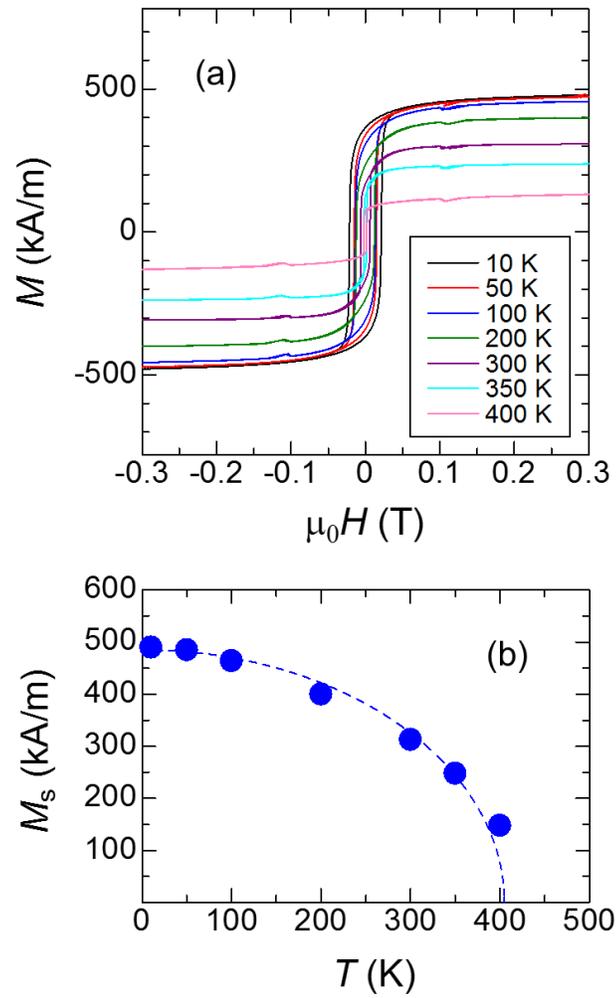

Figure 5: (a) Hysteresis loops of the magnetization $M$ for the films measured with different temperatures $T$. (b) Saturation magnetization $M_s$ as a function of temperature $T$. Dashed curve denotes the calculated data fitted to the experimental data.

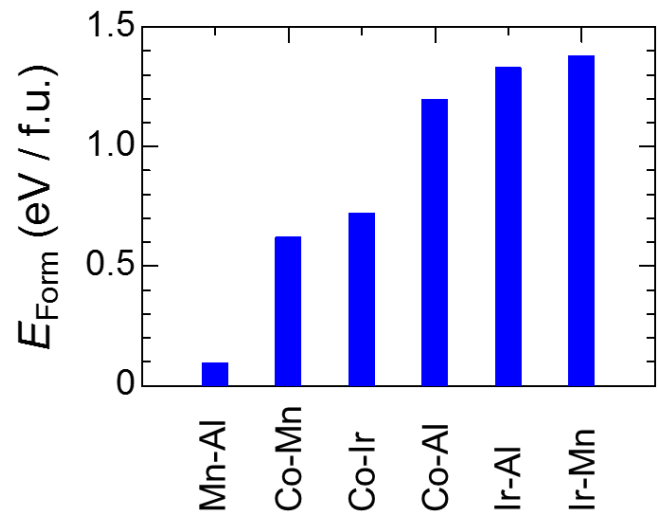

Figure 6: Formation energies $E_{\text{Form}}$ of full-swap disorder of various constituent atoms with respect to that of $Y$ ordered CoIrMnAl, which were evaluated from the first-principles calculations.

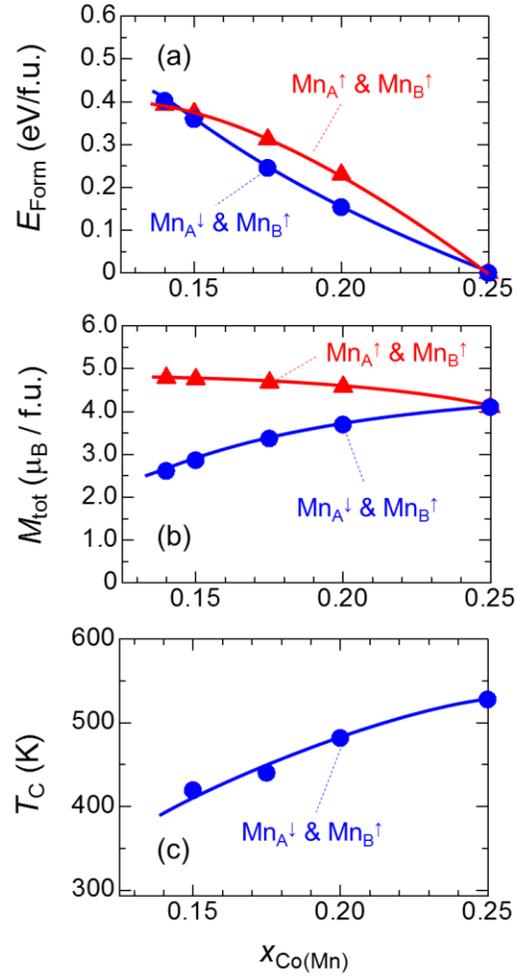

Figure 7: Several properties calculated from the first-principles as a function of the occupation number of Co (Mn) $x_{Co(Mn)}$ at B (A) site in $B2$ CoIrMnAl: (a) Formation energy $E_{Form}$, (b) total magnetic moment $M_{tot}$, and (c) Curie temperature $T_C$. Here, $x_{Co(Mn)} = 0.125$ corresponds to the full-swap disorder of Co-Mn. The $B2$ structure in the inset in Fig. 3(c) with Ir (Al) occupying only at A (B) site was assumed. Circles (triangles) denote the data in case that the spin of Mn at A site is anti-parallel (parallel) to that of Mn at B site. In other word, circles (triangles) denote ferrimagnetic (ferromagnetic) ordering of $B2$ CoIrMnAl. Curves are the visual guides.